\documentclass[aps,preprint,letterpaper,amssymb,showkeys,11pt]{revtex4}

\begin{document}  

\title{The Phase Transition in Statistical Models Defined on Farey Fractions}
\author{Jan Fiala}
\altaffiliation{Department of Physics \& Astronomy, University of Maine, Orono, ME 04469}
\email{jan.fiala@umit.maine.edu}
\author{Peter Kleban}
\altaffiliation{LASST/SERC and Department of Physics \& Astronomy, University of Maine, Orono, ME 04469}
\email{kleban@maine.edu; fax:  (207)-581-2255; phone:  (207)-581-2258}
\author{Ali \"{O}zl\"{u}k}
\altaffiliation{Department of Mathematics and Statistics, University of Maine, Orono, ME 04469}
\email{ozluk@gauss.umemat.maine.edu}
\date{\today}
\begin{abstract} 
We consider several statistical models defined on the Farey fractions. Two of these models may be
regarded as ``spin chains", with long-range interactions, while another arises in the study of multifractals
associated with chaotic maps exhibiting intermittency. We prove that these models all have the same free 
energy. Their thermodynamic behavior is determined by the spectrum
of the transfer operator (Ruelle-Perron-Frobenius operator), which is defined using the maps (presentation functions)
generating the Farey ``tree". The spectrum of this operator was completely determined by Prellberg. It follows that these
models have a second-order phase transition with a specific heat divergence of the form $C \sim [\epsilon\ln^2\epsilon]^{-1}$.
The spin chain models are also rigorously known to have a discontinuity in the magnetization at the phase transition.
\end{abstract}

\maketitle
\section{ Introduction}

In this work we consider several statistical models defined on the Farey fractions.
One is the Farey fraction spin chain, a one-dimensional statistical
 model first proposed by two of the authors \cite{K-O}.  This work has
 spawned a number of further studies, by both physicists and
 number theorists \cite{C-Kn,O,Pe}  One can define the model as
 a periodic chain of sites with two possible spin states (A or B)
 at each site.  The interactions are long-range, which allows a
 phase transition to exist in this one-dimensional system.  The
 Farey spin chain is rigorously known to exhibit a single phase transition
 at temperature $\beta_c = 2$ \cite{K-O}.  The phase transition itself is
 most unusual.  The low temperature state is completely ordered \cite{K-O,C-Kn} .
  In the limit of a long chain, for $\beta > \beta_c$, the
 system is either all A or all B.  Therefore the free energy
 is constant and the magnetization (defined via the difference in the number
 of spins in state A vs. those in state B) is completely saturated over
 this entire temperature range.  Thus, even though the system
 has a phase transition at finite temperature, there are no
 thermal effects at all in the ordered state. 

At temperatures above the phase transition (for $\beta < \beta_c$),
 fluctuations occur, and the free energy decreases with $\beta$.
  Here the system is paramagnetic.  Since there is no symmetry-breaking 
field in the model, the magnetization vanishes.  Thus the magnetization
 jumps from its saturated value in the low temperature phase, to zero
 in the high temperature phase \cite{C-K}.  This might suggest a first-order
 phase transition, but the behavior with temperature is different. In
 this work, we prove that as a function of temperature, the transition
 is second-order, and the same as that which occurs in the Knauf spin chain (see below) and the ``Farey tree" 
multifractal model.  The latter exhibits intermittency, and was studied by Feigenbaum, Procaccia, and T\'{e}l \cite{F}.

The Farey spin chain is defined
 in an unusual way. It is given in terms of the energy of each possible
 configuration, rather than via a Hamiltonian.  There is no known way
 to express the energy exactly in terms of the spin variables \cite{K-O}.  Further,
 numerical results indicate that when one does, the Hamiltonian has all
 possible even interactions (and they are all ferromagnetic), so an 
explicit Hamiltonian representation, even if one could find it, would
 be exceedingly complicated.  

In previous work \cite{K-O}, it was proven that the Farey spin chain free energy (per site,
 in the infinite chain limit)  is the exactly same as the free energy
 of an earlier, related spin chain model due to Knauf \cite{K}.  In the 
present paper, we  extend this result in several ways.  
 
We begin by defining the spin chain and Farey tree models in Section II.
In Section III we prove that the free energy for the Farey tree model is
the same as the free energy of the Knauf model.  This is established
by use of bounds on the Knauf partition function.  In Section IV, we examine
the Farey model, which is specified by the maps (presentation functions \cite{F}) that generate the Farey tree.  The free energy 
in this case is given by the logarithm of the largest eigenvalue $\lambda(\beta)$ 
of the transfer operator. Some years ago, Knauf \cite{K-o} realized that
the free energy of the Knauf model is also given by the logarithm of $\lambda(\beta)$,
(without noting the connection to the Farey tree model, however).
Combining his result with our analysis rigorously shows the equality
of all four free energies-for the Farey spin chain, Knauf model, Farey tree model
and Farey model. In Section V, by using the results of \cite{P-S},
we show that the phase transition is continuous 
(and of second order, i.e. the specific heat is divergent).
It also follows that the phase transition in the Farey model occurs at
the Hausdorff dimension of the Farey tree system, as expected.  We conclude
by briefly pointing out some connections with number theory and mentioning some
implications of scaling theory for the spin chain models.

\section{Definitions}

We use the notation $r_k^{(n)}:=\frac{n_k^{(n)}}{d_k^{(n)}}$ for
 the Farey
 fractions, where
$n$ is the order of the Farey fraction in level $k$. Level $k=0$
 consists
 of the two fractions $\left\{ \frac{0}{1},\frac{1}{1} \right \}$. 
 Succeeding levels are generated by keeping
all the fractions from level $k$ in level $k + 1$, and including new fractions.
  The new fractions at level $k+1$ are defined via
 $ d_{k+1}^{(2n)}:=d_k^{(n)}+d_k^{(n+1)}$ and
 $ n_{k+1}^{(2n)}:=n_k^{(n)}+n_k^{(n+1)}$,
 so that\\
$k=0 \quad  \left \{ \frac{0}{1},\frac{1}{1} \right \}$\\
$k=1 \quad \left \{ \frac{0}{1},\frac{1}{2},\frac{1}{1} \right \}$\\
$k=2 \quad \left \{ \frac{0}{1},\frac{1}{3},\frac{1}{2},\frac{2}{3},
\frac{1}{1} \right \}$, etc.\\[.1cm]
Note that $n=1,\ldots,2^k+1$. When the Farey fractions are defined using matrices
(spin states) A and B, the level $k$ corresponds to the number of matrices and hence the length of the spin chain \cite{K-O}.

It follows that the fractions in a given level are always in increasing order.
 The Farey fractions differ from the Farey ``tree" \cite{F}, where only the
 new fractions are kept at each succeeding level.

The partition function for the Farey spin chain (FC) may be written as \cite{K-O}
\begin{equation}\label{FF}
Z_k^{FC}(\beta):=\sum_{n=1}^{2^k}\frac{1}{(d_k^{(n)}+n_k^{(n+1)})
^{\beta}},\quad \beta
\in \mathbb R.
\end{equation}
Note from (\ref{FF}) that there are $2^k$ states at level $k$
 with energies $E_k^{(n)}= \ln(d_k^{(n)}+n_k^{(n+1)})$. The Farey
 fractions (and hence the energies) can also be defined using
 the spin variables A and B mentioned above \cite{K-O}, but  this
 is not needed here.

For present purposes, it is convenient to use the partition
 function for the Knauf model \cite{C-K}, which is
rigorously known to have the same free energy as the Farey 
spin chain \cite{K-O}.  
The Knauf partition function may be defined via
\begin{equation}\label{a3}
Z_k^K(\beta):=\sum_{n=1}^{2^k}\frac{1}{(d_k^{(n)})^{\beta}},\quad \beta
\in \mathbb R,
\end{equation}
so that a chain of length $k$ has $2^k$ states of energy
$E_k^{(n)}=\ln (d_k^{(n)})$.
The partition function can be written as sum of even and odd terms
\begin{equation}\label{a4}
Z_k^K(\beta)=Z_{k,e}^K(\beta)+Z_{k,o}^K(\beta),
\end{equation}
where
$$Z_{k,e}^K(\beta):=\sum_{n=1}^{2^{k-1}}\frac{1}{(d_k^{(2n)})^{\beta}},
\quad
Z_{k,o}^K(\beta):=\sum_{n=1}^{2^{k-1}}\frac{1}{(d_k^{(2n-1)})^{\beta}}.$$

From the definition of the Farey fractions immediately follows
\begin{equation}\label{a1}
d_{k}^{(2n)}=d_k^{(2n-1)}+d_k^{(2n+1)}
\end{equation}
and
\begin{equation}\label{a2}
d_k^{(2n-1)}=d_{k-1}^{(n)}.
\end{equation}
From (\ref{a1}) we have $$d_k^{(2n)}>d_k^{(2n-1)},\qquad
d_k^{(2n)}>d_k^{(2n+1)},$$
while from (\ref{a2}) we obtain $Z_{k,o}^K(\beta)=Z_{k-1}^K(\beta)$
so that
\begin{equation}\label{a5}
Z_{k,e}^K(\beta)=Z_{k}^K(\beta)-Z_{k-1}^K(\beta).
\end{equation}

The Farey tree model of Feigenbaum, Procaccia and T\'{e}l \cite{F}
 uses the ``Farey tree" rather than the Farey fractions, which 
means retaining only the $2^{k-1}$ even fractions at level $k>1$
so we obtain the set $$\{r_k^{(2n)}|\,n=1,\ldots,2^{k-1}, k>1 \}.$$
The Farey tree partition function is defined by
\begin{equation}\label{a6}
Z_{k}^F(\beta):=\sum_{n=1}^{2^{k-2}}\left(r_k^{(4n)}-r_k^{(4n-2)}\right)
^{\beta}.
\end{equation}
The positive quantities $\left(r_k^{(4n)}-r_k^{(4n-2)} \right)$
 are the radii of the ``balls" in this model. Note that we can also express this partition function using Farey tree denominators only.  One finds
$$Z_{k}^F(\beta)=\sum_{n=1}^{2^{k-2}}\left(\frac{3}{d_k^{(4n)}d_k^{(4n-2)}}\right)
^{\beta}.$$

\section{ Equivalence of the Farey tree and Knauf free energies}

In this section, we show the equivalence of the free energies of the Knauf and Farey tree models.  We begin by finding bounds for the Farey tree partition function $Z^F_k(\beta)$ in terms of the Knauf partition function. We are interested in the case 
$\beta>0$, where there is a phase transition, but it will be easy to see that the free energies are equal for all $\beta \in \mathbb R$.

The Farey fractions satisfy $r_k^{(n)}-r_k^{(n-1)}=1/(d_k^{(n)}d_k^{(n-1)})$. This may be shown for instance using the matrix chain representation in \cite{K-O}. Thus
\begin{eqnarray} \label{a7}
r_k^{(4n)}-r_k^{(4n-2)} & = &
r_k^{(4n)}-r_k^{(4n-1)}+ r_k^{(4n-1)}-r_k^{(4n-2)} \nonumber\\
&=&\frac{1}{d_k^{(4n)}d_k^{(4n-1)}}+\frac{1}{d_k^{(4n-1)}d_k^{(4n-2)}}\\
&>&\frac{1}{\left(d_k^{(4n)}\right)^2},\nonumber
\end{eqnarray}
and similarly $ r_k^{(4n)}-r_k^{(4n-2)} > 1/\left(d_k^{(4n-2)}\right)^2$.
From (\ref{a7}) we also find
\begin{equation}\label{a8}
r_k^{(4n)}-r_k^{(4n-2)}<\frac{2}{\left(d_k^{(4n-1)}\right)^2}.
\end{equation}
Using (\ref{a6}) and (\ref{a7}), for $\beta > 0$, gives
\begin{equation}\label{a9}
Z_{k}^F(\beta)>\sum_{n=1}^{2^{k-2}}\frac{1}{\left(d_k^{(4n)}
\right)^{2\beta}},
\end{equation}
and also $ Z_{k}^F(\beta)>
\sum_{n=1}^{2^{k-2}}1/\left(d_k^{(4n-2)} \right)^{2\beta}$.
Adding these two inequalities we find a lower bound for
the Feigenbaum partition function
\begin{equation}\label{a10}
Z_{k}^F(\beta)>\frac{1}{2}\sum_{n=1}^{2^{k-1}}\frac{1}{\left(d_k^{(2n)}
\right)^{2\beta}}=\frac{1}{2}Z^{K}_{k,e}(2\beta).
\end{equation}
Using the inequality (\ref{a8})
and the relation (\ref{a2}) gives the upper bound
\begin{equation}\label{a11}
Z_{k}^F(\beta)<2^{\beta}\sum_{n=1}^{2^{k-2}}\frac{1}{\left(d_k^{(4n-1)}
\right)^{2\beta}}=2^{\beta}\sum_{n=1}^{2^{k-2}}
\frac{1}{\left(d_{k-1}^{(2n)}
\right)^{2\beta}}=2^{\beta}Z^{K}_{k-1,e}(2\beta).
\end{equation}
Thus the Farey tree partition function at $\beta$ is
bounded both above and below by the even part of the
 Knauf partition function at $2\beta$.
\begin{equation}\label{a12}
\fbox
{$ \displaystyle
\frac{1}{2}Z^K_{k,e}(2\beta)<Z^F_k(\beta)<2^{\beta}Z^K_{k-1,e}(2\beta)
,\quad \beta>0$}
\end{equation}
Similarly, we can find, that
\begin{equation}\label{a12n}
2^{\beta}Z^K_{k-1,e}(2\beta)<Z^F_k(\beta)<\frac{1}{2}Z^K_{k,e}(2\beta)
,\quad \beta<0.
\end{equation}
Finally, for $\beta=0$ it is obvious that
$$Z_k^F(\beta)=\frac{1}{4}Z_k^K(2\beta).$$

The free energy per site is defined by
\begin{equation}\label{freeE}
f(\beta) :=\frac{-1}{\beta} \lim_{k\rightarrow\infty}\frac{\ln
Z_k(\beta)}{k}.
\end{equation}
(Recall that the level $k$ corresponds to the length of the spin chain.)
We now use (\ref{a12}) to prove that
$$f_F(\beta) = f_K(2\beta).$$
where $f_F$ refers to the free energy obtained from $Z_k^F$.

For $\beta>1$ one has \cite{K}
$$Z^K_k(2\beta)\stackrel{k\rightarrow\infty}{\longrightarrow}
\frac{\zeta(2\beta-1)}{\zeta(2\beta)},$$
which implies that $f_K(2\beta)=0$.
Also, by (\ref{a5}),
$$Z^K_{k,e}(2\beta)\stackrel{k\rightarrow\infty}{\longrightarrow}0,$$
and using (\ref{a12}) gives
$$Z^F_k(\beta)\stackrel{k\rightarrow\infty}{\longrightarrow}0.$$
Since $Z^F_k(\beta)>0$,
$$\frac{-\ln Z^F_k(\beta)}{k}\geq 0\Rightarrow f_F(\beta)\geq 0.$$
Note that for $\beta=1$ one has $Z_k^F(1) \leq 1$, since this partition function 
reduces to a simple sum of Farey tree fraction separations (ball lengths), which 
cannot exceed the length of the interval $[0,1]$.  Therefore the inequality 
still holds (and in fact, as shown below, $f_F(1)=0$).

Now clearly
$$Z^K_{k,e}(2\beta)>\frac{1}{(k+1)^{2\beta}}$$
so by (\ref{a12}) we find
$$Z^F_k(\beta)>\frac{1}{2}\frac{1}{(k+1)^{2\beta}},$$
and
$$0\leq\frac{-\ln Z^F_k(\beta)}{k}<\frac{2\beta\ln(k+1)}{k}+\frac{\ln
2}{k}.$$
Thus we have
\begin{equation}\label{betag}
\fbox{$\displaystyle f_F(\beta) = f_K(2\beta)=0\ {\rm for}\ \beta \geq 1.$}
\end{equation}
The validity of  $f_K(2)=0$ is clear from the treatment in \cite{K-o} and the remark at the end
of this section.

For $\beta < 1$ we can write
$$Z^K_{k,e} = Z^K_k-Z^K_{k-1} =
Z^K_k\left(1-\frac{Z^K_{k-1}}{Z^K_{k}}\right),$$
so
\begin{equation}\label{eq}
-\frac{\ln Z^K_{k,e}}{k} = -\frac{\ln
Z^K_k}{k}-\frac{\ln\left(1-\frac{Z^K_{k-1}}{Z^K_{k}}\right)}{k}.
\end{equation}
It is shown in \cite{C-K} (by arguments using the transfer operator, see below) that for $0<\beta<1$
the free energies obtained from $Z_k^K$ and $Z_{k,e}^K$ are the same, thus 
for $k\rightarrow\infty$
\begin{equation}\label{term}
\frac{\ln\left(1-\frac{Z^K_{k-1}}{Z^K_{k}}\right)}{k}\rightarrow 0.
\end{equation}
(This also can be shown directly by considering the equation $Z_k^K(2\beta)=1+\sum_{j=1}^{k}Z_{j,e^K}(2\beta)$,
which follows from (\ref{a5}). For  $0<\beta<1$ the series is bounded by a geometric series because of the 
inequality $Z_{k,e}^K>2^{1-\beta}Z_{k-1,e}^K.$)
For $\beta \le 0$ it is easy to check that $Z_{k-1,e}^K(2\beta) /Z_{k,e}^K(2\beta) \le 1/2$. Thus (\ref{term}) holds for
all $\beta <1$.

 Using  (\ref{a12}) (and, for $\beta \leq 0$, (\ref{a12n}) and the line below) then
establishes 
\begin{equation}\label{betal}
\fbox{$\displaystyle f_F(\beta) = f_K(2\beta)\ {\rm for}\ \beta < 1.$}
\end{equation}

Note that, as mentioned, the Knauf partition function $Z^K_k (2\beta)$ is finite as $k \to \infty$ for $\beta > 1$ \cite{K}. 
Using (\ref{a5}) and (\ref{a12}) one sees immediately that the Farey tree partition function $Z^F_k (\beta)$ vanishes in 
this limit for $\beta > 1$. At $\beta = 1$, it follows immediately from the definition (\ref{a6}) and simple properties of
 the Farey fractions that $0 < Z^F_k (1) <1$. For $\beta <1$, since $f_K(2 \beta) < 0 $ \cite{C-K} and using (\ref{betal})
 and (\ref{freeE})  it follows that $Z^F_k (\beta)$ is infinite. This establishes rigorously that the Hausdorff dimension
 of the set formed by the ``balls" is $ \beta_H = 1$, as expected.

Finaly, consider (\ref{a12}) and the fact, mentioned above, that $ Z^F_k (1) <1$. It follows that
$$Z_{k,e}^K(2)=\sum_{n=1}^{2^{k-1}}\frac{1}{(d_k^{(2n)})^2}<2,$$
so that this sum over the ``new" Farey denominators is bounded by $2$ at all levels. Since the 
``new" denominators at level $k-1$ become ``old" denominators at level $k$, one also sees that $Z_{k}^K(2)\le 2k+1$.

\section{Transfer operator approach}

In this section we consider the transfer operator (Ruelle-Perron-Frobenius operator) of the Farey map. 
The previous section shows rigorously that the free energies of the Knauf and Farey fraction spin chain and Farey tree model are the same.  Here we prove that they (as well as the free energy of the Farey tree model in a certain approximation specified below) are simply given by the largest eigenvalue of this operator.  The next section considers the asymptotic behavior of this eigenvalue near the phase transition, known from the work of Prellberg \cite{P-s}, which specifies the order of the phase transition.

The Ruelle-Perron-Frobenius operator $\cal K$
associated with a map $f$ (piecewise monotonic transformation of closed interval $I$) is given by 
\begin{equation}\label{RPF}
{\cal K}_{\beta}\varphi(x)=\sum_{f(y)=x}|f'(y)|^{-\beta}\varphi(y),\quad \beta \in \mathbb R,
\end{equation}
where the sum is over each strictly monotonic and continuous piece of $f$ satisfying
the summation condition. See \cite{P-S,Diss} for a more complete discussion.

The Farey map is defined by \cite{F,P-S}
\begin{equation}\label{FM}
f(x)=
\left\{ \begin{array}{ll}
			f_0(x)=x/(1-x), & 0\le x \le 1/2, \\
			f_1(x)= (1-x)/x,& 1/2<x\le 1.
		\end{array} \right. 
\end{equation}	
The operator then consists of two corresponding terms ${\cal K}_0$ and ${\cal K}_1$ which can be identified as 
``intermittent" and ``chaotic" parts, respectively \cite{P-s}. We may write ${\cal K}_{\beta}={\cal K}_0+{\cal K}_1$
where ${\cal K}_i\varphi(x)=|F'_i(x)|^{\beta}\varphi(F_i(x))$ and the ``presentation function" 
\cite{F} $F_i$ is the inverse map of $f_i$ (see (\ref{pre}) below). Thus
\begin{equation}\label{fo}
{\cal K}_{\beta}\varphi(x)=(1+x)^{-2\beta}
\left[\varphi\left(\frac{x}{1+x}\right)+
\varphi\left(\frac{1}{1+x}\right)\right], \quad \beta \in \mathbb R.
\end{equation}
Following the thermodynamic formalism approach \cite{R} 
it was shown in \cite{P-S,Diss} that the largest eigenvalue of ${\cal K}_{\beta}$ in (\ref{fo}) (defined on the space of functions
with bounded variation) is related to a free energy via $f(\beta)=-\beta ^{-1}\ln\lambda(\beta)$ for $\beta \in \mathbb R$.
We call this the free energy of the Farey model. 

In this section we consider ${\cal K}_{\beta}$ acting on $L^2$ and show that the free energy obtained from its largest eigenvalue is the same as the free energy of the Knauf and Farey tree model (in its original version or using the approximation below) for $0 < \beta < 1$.  In the next section, we prove that the free energy of the Farey model in this $\beta$ range is also the same. For $\beta > 1$, the free energy of any of these models is already known to be zero (see section III or \cite{P-s}).

The Knauf spin chain at level $k-1$ may be described by a vector $Y_{k-1}(2\beta) \in l^2(\mathbb N_0)$, the first component of which is the ``even" Knauf
partition function $Z^K_{k,e}(2\beta)$. The ``transfer operator" of the Knauf spin chain then maps $Y_{k-1}(2\beta)$ to the next level:
\begin{equation}\label{recuy}
Y_k(2\beta)=\tilde{\cal C}(2\beta)Y_{k-1}(2\beta),
\end{equation}
where $\tilde{{\cal C}}(2\beta):l^2(\mathbb N_0)\rightarrow l^2(\mathbb N_0)$ and \cite{C-K}
\begin{widetext}
\begin{equation}\label{Co}
\tilde{C}(2\beta)_{i,j}=(-1)^j\,2^{-2\beta-i-j}\left [\left(
\begin{array}{c}
 -2\beta -i\\
 j
\end{array}
\right )+
\sum_{s=0}^{i}2^s
\left (
\begin{array}{c}
 i\\
 s
\end{array}\right)
\left (
\begin{array}{c}
 -2\beta -i\\
 j-s
\end{array}\right)
\right ],
\end{equation}
\end{widetext}
$(i,j \in \mathbb N_0)$, with the generalized binomial coefficients
$\left (\begin{array}{c} a\\ b
\end{array}\right)=(\Pi_{i=1}^{b-1}(a-i))/b!$, $a\in \mathbb R$,
$b\in \mathbb N_0$, and $\left (\begin{array}{c} a\\ b
\end{array}\right)=0$ if $b<0$.
Knauf \cite{K-o} has further shown that for $0<\beta <1$, $\tilde{{\cal C}}(2\beta)$ has the same largest eigenvalue $\lambda (\beta)$ 
as ${\cal K}_{\beta}:L^2((0,1)) \rightarrow L^2((0,1))$. The argument involves expanding (\ref{fo}) about $x=1$ with $\varphi(x)=\sum_{m=0}^{\infty} a_m (1-x)^m$.  Doing this, one finds that the action of ${\cal K}_{\beta}$ on the quantities $a_m$ (note that $a_m = (-1)^m\varphi^{(m)}(1)/{m!}$) is given by $\tilde{\cal C}^T(2\beta)$, where $T$ denotes transpose.

In addition, $\tilde{\cal C}^T(2\beta)$ is independent of $k$, so the components of the vector $X_k(2\beta)$ (defined using (\ref{recuy}) with $\tilde{\cal C}^T(2\beta)$ replacing $\tilde{\cal C}(2\beta)$) are proportional to
the Taylor series coefficients of an associated function $\phi^{(\beta)}_{k}(x)$. This function therefore satisfies
\begin{equation}\label{fr}
\phi_k^{(\beta)}(x)=(1+x)^{-2\beta}
\left[\phi_{k-1}^{(\beta)}\left(\frac{x}{1+x}\right)+
\phi_{k-1}^{(\beta)}\left(\frac{1}{1+x}\right)\right].
\end{equation}

It is shown in \cite{C-K} that $\tilde{\cal C}(2\beta)$ (and hence $\tilde{\cal C}^T(2\beta)$) is
an operator of Perron-Frobenius type for $0<\beta <1 $.
Thus $\lambda (\beta)$ is a simple eigenvalue (the same for $\tilde{\cal C}$ 
or $\tilde{\cal C}^T$).  The corresponding eigenvector is strictly
positive and unique, and may be obtained (for $\tilde{\cal C}^T$) via $V(2\beta)=\lim_{k\to \infty}X_k(2\beta)/||X_k(2\beta)||$. In addition, it follows
that for $0<\beta<1$ the eigenvalue $\lambda (\beta)>1$
is an analytic function of $\beta$, and its positive normalized
eigenvector $V(2\beta)$ is analytic in $\beta$.
Hence
\begin{equation}\label{ps}
\phi_k^{(\beta)}\sim \lambda(\beta)^k \phi^{(\beta)},
\end{equation}
where $\phi^{(\beta)}(x)$ is the normalized eigenvector of ${\cal K}_{\beta}:L^2((0,1)) \rightarrow L^2((0,1))$ corresponding to $V(2 \beta)$. 
Substituting this result in (\ref{fr}) we get, for $0<\beta <1$,
\begin{equation}\label{f}
\lambda (\beta)\phi^{(\beta)}(x)=(1+x)^{-2\beta}
\left[\phi^{(\beta)}\left(\frac{x}{1+x}\right)+
\phi^{(\beta)}\left(\frac{1}{1+x}\right)\right],
\end{equation}
which is equivalent to (\ref{fo}) when $\lambda (\beta)$ is
the maximal eigenvalue. 
Then 
\begin{equation}\label{eigen}
\lim_{k\to \infty}\frac{Z^K_{k,e}(2\beta)}{Z^K_{k-1,e}(2\beta)}=\lambda (\beta)
\end{equation}
together with (\ref{freeE}), (\ref{eq}) and (\ref{term}) give us the Knauf free energy as expected
\begin{equation}\label{freeK}
f_K(2\beta)=-\frac{1}{\beta}\ln\lambda (\beta),\quad 0<\beta <1.
\end{equation}
Note that for $\beta \ge 1$, $f_K(2\beta)=0$ (see section III) and also that $f(\beta)=0$ for $\beta \ge 1$ follows from the spectrum
of the operator ${\cal K}_{\beta}$ (\cite{P-s}, see also the next section). Thus the free energy of the Farey spin chain, Farey tree and Knauf
models are given by the largest eigenvalue of the Ruelle-Perron-Frobenius operator for $\beta >0$. 

To further examine these connections we follow the treatment in \cite{F}.
We focus on (\ref{f}) and make use of presentation functions. The Farey tree
can be generated by two presentation functions
\begin{equation}\label{pre}
F_0=\frac{x}{1+x},\qquad F_1=1-F_0=\frac{1}{1+x}.
\end{equation}
Every fraction at each level $k>1$ of the Farey tree can be reached by composition of
$k$ functions $F_{\epsilon}$ ($\epsilon \in \{0,1\}$) evaluated at
$x^*=\frac{1}{2}$. For example, at level $k=3$,  $F_0\circ F_1(\frac{1}{2})=\frac{2}{5}=r_3^{(4)}$.
So the diameter of every ``ball" in the Farey tree model (see (\ref{a6})) can be written as
\begin{equation}\label{ball}
r_k^{(4n)}-r_k^{(4n-2)} =|F_{\epsilon _1}\circ F_{\epsilon
_2}\circ \ldots \circ F_{\epsilon _{k-1}}(F_0(x^*))
-F_{\epsilon _1}\circ F_{\epsilon
_2}\circ \ldots \circ F_{\epsilon _{k-1}}(F_1(x^*))|.
\end{equation}
Note that the sequence of presentation functions in the two
 Farey fractions in (\ref{ball}) is identical except for the
 $F_{\epsilon_k}$, i.e. only the presentation functions applied first to $x^*$ differ.
As $k \rightarrow \infty$, the diameter of the balls converges 
to zero (this follows easily from (\ref{a7})). Therefore it is
reasonable to suppose that for $k$ sufficiently large each 
diameter can be approximated by the derivative of the composed
 function with respect to $x^*$. Then, using the chain rule, (\ref{ball})
behaves asymptotically as
\begin{equation}\label{assym}
r_k^{(4n)}-r_k^{(4n-2)}\sim |F'_{\epsilon _1}(F_{\epsilon
_2}\circ F_{\epsilon _3}\circ \ldots)
F'_{\epsilon _2}(F_{\epsilon
_3}\circ F_{\epsilon _4}\circ \ldots)\ldots|.
\end{equation}
Thus we can write for the partition function
\begin{equation}\label{partition}
Z_k^F\sim \ldots \sum_{\epsilon_k}|F'_{\epsilon _k}(F_{\epsilon
_{k+1}}\circ F_{\epsilon _{k+2}}\circ \ldots)|^{\beta}
\sum_{\epsilon_{k-1}}|F'_{\epsilon _{k-1}}(F_{\epsilon
_{k}}\circ F_{\epsilon _{k+1}}\circ \ldots)|^{\beta}
\ldots
\end{equation}
Notice that the sum over $\epsilon_{k}$ and all lower indexed
sums to its right depend only upon
$(F_{\epsilon_{k+1}}\circ F_{\epsilon _{k+2}}\circ \ldots)$. This motivates
 the definition 
\begin{equation} \label{psi}
\psi_{k}^{(\beta)}(x)
:= \sum_{\epsilon_k}|F'_{\epsilon _k}(x)|^{\beta}
\sum_{\epsilon_{k-1}}|F'_{\epsilon _{k-1}}(F_{\epsilon
_{k}}(x))|^{\beta}
\ldots , 
\end{equation}
where $(F_{\epsilon_{k}}\circ F_{\epsilon _{k+1}}\circ \ldots)$ is
denoted by $x$.
One then finds
\begin{equation}\label{recu}
\psi_{k}^{(\beta)}(x)
=\sum_{\epsilon}|F'_{\epsilon}(x)|^{\beta}
\psi_{k-1}^{(\beta)}(F_{\epsilon}(x)).
\end{equation}
Note that since each presentation function $F_{\epsilon}$ 
is a ratio of polynomials, one can extend the
definition of $\psi_k^{(\beta)}(x)$ to the whole interval $[0,1]$. 
Substituting for $F$ and $F'$ we obtain (\ref{fr}) (with $\psi_k$ replacing $\phi_k$).
Therefore choosing $\psi_0^{(\beta)}(x) > 0$ we find 
$\psi_k^{(\beta)} \to \psi^{(\beta)}$ as $k\to \infty$, with the
function $\psi^{(\beta)}$ proportional to $\phi^{(\beta)}$ (the eigenfunction with
the maximum eigenvalue $\lambda(\beta)$). 
This establishes that the approximation (\ref{assym}) is exact
in the limit $k\to \infty$, as expected.

Finally, it is interesting to note some connections with number theory. Specifically, for $\lambda =1$, (\ref{f})
is known as the Lewis
equation and has been studied (for complex $\beta$) because of its connection to the Selberg $\zeta$-function and period polynomials (cusp forms of the modular group) \cite{L-Z}.  
An operator related to ${\cal K}_{\beta}$ (\ref{fo}) also appears in this context and is called the Mayer operator \cite{M}.

\section{ Order of the phase transition and discussion}
In the preceding, we have shown that the Farey spin chain \cite{K-O}, the Knauf spin chain \cite{C-K} and (either version of) the Farey tree model \cite{F} all have the same free energy.  Further, for $0 < \beta < 1$ their free energy is given by the largest eigenvalue of the Farey model transfer operator acting on $L^2$ (\ref{fo}).  Here we show that the transfer operator acting on the space of functions of bounded variation has the same leading eigenvalue in this $\beta$ range, which allows us to make use of the results of Prellberg.  The corresponding equality of free energies for $\beta > 1$ (where the free energy vanishes) follows from known results, as remarked in the previous section.

Prellberg has examined the spectrum of this operator acting on the space of functions with bounded variation  \cite{P-S} (details are in \cite{Diss}).  In order to make use of his results, we must show that the largest eigenvalue in this space is the same as that in $L^2((0,1))$.  To prove this we examine the corresponding eigenvectors.  Expanding $\varphi(x)$, the eigenvector in the $L^2$ space, about $x=1$ as above, one has $\varphi(x)=\sum_{m=0}^{\infty} a_m (1-x)^m$. Thus $\varphi(1)$ is finite, since the coefficients $a_m$ in this expansion are proportional to the components of the eigenvector of $\tilde{C}^T(2 \beta)$ of largest eigenvalue (see section IV).  Furthermore, the $a_m$ are all positive, since the eigenvector of $\tilde{C}^T$ is positive.  Therefore $\varphi(x)$ is a (strictly) decreasing function on $[0,1]$.  Finally, setting $x=0$ in (\ref{f}) shows that $\varphi(0)$ is finite whenever $\lambda \ne 1$. Therefore, $\varphi(x)$ is of bounded variation for $0 < \beta <1$, and since both eigenvectors are unique (up to multiplicative constants) their eigenvalues must coincide in this range of $\beta$ values.

The result of Prellberg of interest here is $$\beta f(\beta)=c\frac{1-\beta}{\ln (1-\beta)}\left [ 1+o(1)\right ],\quad 0<\beta<1,$$
where $c > 0$, and $\beta f(\beta)=0$ for $\beta \geq 1$. This form for the free energy is equivalent to that given in \cite{F}, as may be seen 
by use of the Lambert $W$-function.

The non-analyticity at $\beta =1$ results in a phase transition of second order, since the second derivative of $f(\beta)$ diverges as $\Bigl [(1-\beta)(\ln(1-\beta))^2\Bigr ]^{-1}$ as $\beta \rightarrow 1^-$. This result agrees with \cite{C-K}, where it is proven rigorously that the phase transition is at most second order. Note that the largest eigenvalue is discrete for $\beta <1$. For $\beta > 1$, the discrete spectrum disappears and the largest eigenvalue becomes $\lambda = 1$, which is the upper boundary of the continuous spectrum for all $\beta$.

Our result for the free energy also has some implications for the number of states of the spin chain models. The Knauf
model partition function may be expressed as a Dirichlet series \cite{K}
\begin{equation}\label{D}
Z_k^K(\beta)=\sum_{n=1}^{\infty}\phi_k(n)n^{-\beta},
\end{equation}
where $\phi_k(n)$ is non-zero when $n$ is a Farey denominator at level $k$. This function converges from below
to the Euler totient function $\phi(n)$ as $k\to \infty$. Since the energy of an allowed state is $E=\ln n$, $\phi_k(n)$
gives the number of states of energy $E$ at level $k$.  The functions $\phi_k$ and $\phi$ are very irregular. Our result for the free
energy then shows how the Dirichlet series in (\ref{D}) diverges as $k\to \infty$ for small (but positive!) $(2-\beta)$. (Recall that the phase transition in the spin chains appears at $\beta_c = 2$, since a factor of $2$ appears in comparing with the Farey tree model, see (\ref{freeK}).)
For the Farey spin chain, an equation with the same form as (\ref{D}) may also be written,  with the same leading divergent behavior.
Here the limit of the function corresponding to $\phi_k(n)$ is not known, though some related information is available \cite{Pe}.

One can also consider the implications of scaling theory for the two spin chain models.  It is known that the
 magnetization (defined via the difference in the number of spins in state A vs. those in state B) is one for
 temperatures below the transition and zero above it \cite{C-Kn,C-K}.  Thus the magnetization jumps from its
 fully saturated value to zero at the transition.  This would lead one to suspect a first-order transition,
 but as we have seen, the behavior with temperature is second-order.  However, both these results seem to
 be consistent with scaling theory, with renormalization group eigenvalues $y_T = d$ and $y_h = d$,
 where $d$ is the dimensionality, and using
$(2-\beta)/\ln (2-\beta)$ as the temperature scaling variable.  We plan to report more fully on this elsewhere.
\begin{acknowledgments}
We thank T. Prellberg for providing a copy of \cite{Diss}. Useful conversations with D. Bradley and Don Zagier are gratefully acknowledged.

This research is based on work supported in part by the National Science Foundation under Grant No. DMR - 0203589.
\end{acknowledgments}


\begin{thebibliography}{15}

\bibitem {K-O}
P. Kleban, and \"{O}zl\"{u}k, {\it A Farey fraction spin chain}, Commun. Math. Phys. {\bf 203}, 635-647 (1999).
\bibitem{O} J. Kallies, A. \"Ozl\"uk, M. Peter and C. Snyder,{\it On asymptotic properties of a number
 theoretic function arising from a problem in statistical mechanics} Commun. Math. Phys. {\bf 222}, 9-43 (2001). 
\bibitem{Pe}  M. Peter, {\it The limit distribution of a number-theoretic function arising from
 a problem in statistical mechanics}, J. Number Theory {\bf 90}, 265-280 (2001).
\bibitem{C-Kn} P. Contucci, P. Kleban, and A. Knauf,{\it A fully magnetizing
 phase transition}, J. Stat. Phys. {\bf 97} 523-539 (1999).
\bibitem{C-K} P. Contucci, and A. Knauf, {\it The phase transition of the number-theoretic
 spin chain}, Forum Mathematicum {\bf 9}, 547-567 (1997).
\bibitem{F} Feigenbaum, M. J., Procaccia, and T. Tel, 
{\it Scaling properties of multifractals as an eigenvalue problem}, Phys. Rev. A {\bf 39}, 5359-5372  (1989).
\bibitem{K} A. Knauf, {\it On a ferromagnetic spin chain}, Commun. Math. Phys. {\bf 153}, 77-115 (1993).
\bibitem{K-o}A. Knauf, {\it The number-theoretical spin chain and the Riemann zeros},
 Commun. Math. Phys. {\bf 196}, 703-731  (1998).
\bibitem{P-S} T. Prellberg, and J. Slawny, 
{\it Maps of intervals with indifferent fixed points: 
thermodynamic formalism and phase transition}, J. Stat. Phys. {\bf 66}, 503-514 (1992).
\bibitem{P-s} T. Prellberg, 
{\it Complete determination of the spectrum of a transfer operator associated with intermittency}, 
preprint [arXiv: nlin.CD/0108044], (2001).
\bibitem{Diss} T. Prellberg, {\it Maps of intervals with indifferent fixed points: 
thermodynamic formalism and phase transition}, Ph.D. thesis, Virginia Tech (1991).
\bibitem{R} D. Ruelle, {\it Thermodynamic Formalism}, Addison-Wesley, (1978).
\bibitem{L-Z} J. Lewis and D. Zagier, {\it Period functions for Maass wave forms}, 
Annals of Mathematics {\bf 153}, 191-258 (2001).
\bibitem{M} D. Mayer, {\it The thermodynamic formalism approach to Selberg's zeta 
function for PSL(2,$\mathbb Z$)}, Bull. AMS {\bf 25}, 55-60 (1991).



\end{thebibliography}
\end{document}